 \documentclass[sigplan,screen]{acmart}
\AtBeginDocument{%
  }

\setcopyright{acmlicensed}
\copyrightyear{2018}
\acmYear{2018}
\acmDOI{XXXXXXX.XXXXXXX}
\acmConference[Conference acronym 'XX]{Make sure to enter the correct
  conference title from your rights confirmation email}{June 03--05,
  2018}{Woodstock, NY}
\acmISBN{978-1-4503-XXXX-X/2018/06}

\usepackage{todonotes}

\usepackage{algorithm}
\usepackage{algpseudocode}
\usepackage{amsmath}
\usepackage{array}
\usepackage{listings}
\usepackage{booktabs, multirow}
\usepackage{pifont}

\usepackage{tikz}
\usetikzlibrary{arrows.meta, positioning, fit, backgrounds, calc}
\definecolor{sysmc}{HTML}{DCEEFB}
\definecolor{syshc}{HTML}{FFF3E0}
\definecolor{syskc}{HTML}{E3F2E6}
\definecolor{sysdc}{HTML}{F0E4F5}
\definecolor{syskci}{HTML}{C8E6C9}

\newcommand{\tool}{SRDatalog}
\newcommand{\cmark}{\ding{51}}
\newcommand{\xmark}{\ding{55}}

\begin{document}

\title{Scaling Worst-Case Optimal Datalog to GPUs}


\author{Yihao Sun}
\email{ysun67@syr.edu}
\affiliation{%
  \institution{Syracuse University}
  \city{Syracuse}
  \state{New York}
  \country{USA}
}

\author{Kunting Qi}
\email{kqi@uic.edu}
\affiliation{%
  \institution{University of Illinois, Chicago}
  \city{Chicago}
  \state{Illinois}
  \country{USA}
}

\author{Thomas Gilray}
\email{thomas.gilray@wsu.edu}
\affiliation{%
  \institution{Washington State University}
  \city{Pullman}
  \state{Washington}
  \country{USA}
}

\author{Sidharth Kumar}
\email{sidharth@uic.edu}
\affiliation{%
  \institution{University of Illinois, Chicago}
  \city{Chicago}
  \state{Illinois}
  \country{USA}
}

\author{Kristopher Micinski}
\email{kkmicins@syr.edu}
\affiliation{%
  \institution{Syracuse University}
  \city{Syracuse}
  \state{New York}
  \country{USA}
}

\renewcommand{\shortauthors}{Sun et al.}


\begin{abstract}
Datalog is a declarative logic-programming language used for complex analytic reasoning workloads such as program analysis and graph analytics. Datalog's popularity is due to its unique price-point, marrying logic-defined specification with the potential for massive data parallelism. While traditional engines are CPU-based, the memory-bound nature of Datalog has led to increasing interest in leveraging GPUs. These engines beat CPU-based engines by operationalizing iterated relational joins via SIMT-friendly join algorithms.

Unfortunately, all existing GPU Datalog engines are built on binary joins, which are inadequate for the complex multi-way queries arising in production systems such as DOOP and \textsf{ddisasm}. For these queries, binary decomposition can incur the AGM bound asymptotic blowup in time and space, leading to OOM failures regardless of join order. Worst-Case Optimal Joins (WCOJ) avoid this blowup, but their attribute-at-a-time intersections map poorly to SIMT hardware under key skew, causing severe load imbalance across Streaming Multiprocessors (SMs). We present \tool{}, the first GPU Datalog engine based on WCOJ. \tool{} uses flat columnar storage and two-phase deterministic memory allocation to avoid the OOM failures of binary joins and the index-rebuild overheads of static WCOJ systems. To mitigate skew and hide hardware stalls, \tool{} further employs root-level histogram-guided load balancing, structural helper-relation splitting, and stream-aligned rule multiplexing. On real-world program-analysis workloads, \tool{} achieves geometric-mean speedups of 21$\times$ to 47$\times$. 

\end{abstract}


\received{20 February 2007}
\received[revised]{12 March 2009}
\received[accepted]{5 June 2009}

\maketitle

\section{Introduction}


Datalog has become a ubiquitous part of analytic reasoning, with applications in graph mining~\cite{seo2013socialite}, program analysis~\cite{whaley2005using,scholz2016fast,ptanalysis}, binary reverse engineering~\cite{flores2020datalog}, and neuro-symbolic reasoning~\cite{li2023scallop,lobster}. While many state-of-the-art Datalog engines target CPUs (e.g., Souffl\'e~\cite{jordan2016souffle} and FlowLog~\cite{zhao2025flowlog}), the massive potential data parallelism inherent to many Datalog applications has motivated the recent development of GPU-based Datalog engines~\cite{sun2025column,sun2025optimizing,lobster}. 

Datalog engines work by recursively executing a set of \emph{rules} to a fixed point. Each rule is written as a Horn clause (see \S\ref{sec:bg-datalog}), which deduces (and ``materializes'') a new fact (tuple) whenever some conjunction holds. Because fixed-point materialization forms the bulk of Datalog's runtime, most modern Datalog engines compile these rules to relational algebra kernels~\cite{jordan2016souffle,nenov2015rdfox,sahebolamri2022seamless, zhao2025flowlog}; this approach  offers a clean separation between the program's semantics and the underlying implementation of the relational algebra. In practice, the vast bulk of Datalog's execution time is spent doing large relational joins~\cite{abiteboul1995foundations}. Initially, many Datalog engines used pipelined binary join, breaking down a multi-way join $R_0 \bowtie R_1  \bowtie R_2 \bowtie \cdots $ into a series of binary joins $((R_0 \bowtie R_1) \bowtie R_2) \bowtie \cdots $ and then realizing each subexpression via range-indexed nested-loop joins~\cite{jordan2016souffle}. However, for cyclic queries (\emph{e}.\emph{g}., triangle counting), the AGM bound ensures that \emph{every} binary join is destined to asymptotic failure, in the sense that materializing the results of intermediate sub-queries incurs superlinear work compared to the cardinality of the output~\cite{agm}. 

To cope with the asymptotic overhead of poor binary join plans, \emph{worst-case optimal joins} (WCOJs) have emerged as a potential solution~\cite{ngo2018worst}. These join algorithms avoid the intermediate materialization (either temporal or spatial) by doing the join one variable at a time, effectively deduplicating redundant work due to ``dangling tuples'' which are enumerated and then filtered out in operator-at-a-time joins such as pipelined binary join~\cite{agm}. While the asymptotic guarantees of worst-case optimal joins are attractive, conventional wisdom initially dictated that their constant factors and engineering complexity made them unsuitable for Datalog. However, recent graph-analytics engines such as HoneyComb~\cite{wu2025honeycomb} and cuMatch~\cite{park2025cumatch} have demonstrated the success of WCOJ in practice for massive, multi-way join processing. Along with worst-case optimality, WCOJs also offer a degree of \emph{robustness}, avoiding the wildly-unstable performance for large programs. For example, the DOOP program analysis contains numerous rules which include 6--8 conjoined literals: in Souffl\'e, such rules tend to be deeply unstable, with their performance varying by orders of magnitude depending on the binary plan chosen. By contrast, WCOJs ensure reasonably-good complexity (albeit not \emph{instance}-optimal) regardless of plan particulars (variable ordering)~\cite{wang2023adopt}.



We present the first-ever approach to WCOJs on modern SIMT GPUs.
Alongside the engineering aspects of putting WCOJ on the GPU, our work contends with a fundamental mechanical vulnerability: WCOJ's attribute-at-a-time intersection model is uniquely susceptible to structural data skew~\cite{ngo2014skew}. For example, in program analysis, when an inner join variable binds to a ubiquitous high-degree hub (e.g., this occurs in the DOOP points-to analysis for Java when dispatching on \textsf{java.lang.Object}), the evaluating thread group is trapped processing a massive, localized intersection space. This induces severe macroscopic load imbalance; the majority of the GPU's thread groups finish their partitions quickly and sit idle, waiting on a single thread group bottlenecked by a disproportionately large workload. Existing systems attempt to resolve this load imbalance, but introduce their own structural flaws: static partitioned systems (e.g., cuMatch~\cite{park2025cumatch}) rely on hierarchical indices whose continuous reconstruction bottlenecks Datalog's delta-merge phase, while dynamic systems (e.g., ALFTJ~\cite{lai2022accelerating}) manage skew via runtime lock-free queues, introducing SIMT-hostile atomic serialization during write-intensive workloads.
    


We introduce \tool{}, the first GPU-accelerated WCOJ Datalog engine built to resolve the tension between algorithmic robustness and strict physical hardware constraints. Rather than relying on heavyweight hierarchical indices or dynamic work-stealing queues, \tool{} adapts WCOJ to execute strictly within SIMT-sympathetic boundaries: it utilizes flat columnar storage, deterministic two-phase memory allocation, and lightweight launch-time skew scheduling. By refusing to compromise on these core memory pipelines, \tool{} delivers a robust, high-performance execution foundation for complex, production-grade Datalog programs.

%

In summary, this paper makes four core contributions:
\begin{itemize}
    \item We introduce an \emph{iterative GPU WCOJ architecture} that strictly enforces flat columnar storage and two-phase deterministic memory allocation, avoiding the intermediate memory explosion of binary joins and the index-rebuild overhead of static systems.
    \item We resolve the inherent load imbalance of large multi-way joins via \emph{root-level histogram load balancing} that deterministically resolves outermost skew at kernel launch, combined with \emph{structural helper-relation splitting}, mitigating imbalance.
    \item An approach to stream-parallel rule scheduling, leveraging the CALM theorem~\cite{hellerstein2020keeping} to overlap independent compute kernels and memory allocation stalls across concurrent CUDA streams, sustaining high SM occupancy on fractured Datalog strata.
    \item \tool{}, the fastest and most scalable GPU Datalog engine to date. 
We comprehensively evaluate \tool{} against five state-of-the-art systems on real-world program analysis workloads. \tool{} demonstrates geometric mean speedups of 21$\times$ to 47$\times$ over cost-equivalent CPU baselines, avoids the intermediate memory exhaustion of binary-join GPU engines, and outperforms static WCOJ accelerators by up to 4.0$\times$.
\end{itemize}

\section{Background}
\paragraph*{Datalog} \label{sec:bg-datalog}
Datalog is a declarative logic language originally designed for deductive databases~\cite{abiteboul1995foundations} and now widely applied to data-intensive domains including program analysis~\cite{scholz2016fast}, binary reverse engineering~\cite{flores2020datalog}, and neuro-symbolic reasoning~\cite{li2023scallop}.
A Datalog program consists of a set of definite Horn-clauses:
\[
\textit{Head}(\ldots) \;\leftarrow\;
\textit{Body}_1(\ldots),\;\ldots,\;\textit{Body }_n(\ldots).
\]
The head specifies a tuple to be derived, and the body specifies the conjunction of conditions under which the head tuple can be derived. Unlike other database query languages, Datalog is particularly well suited to recursive queries. In a recursive Datalog program, the same relation can appear in both the head and the body of a rule. For example, the classic Transitive Closure (TC) program recursively computes reachability over a graph:
\[
\begin{array}{lcl}
\textit{TC}(x, y) & \leftarrow & \textit{Edge}(x, y), \\
\textit{TC}(x, y) & \leftarrow & \textit{TC}(x, z),\;
                                 \textit{Edge}(z, y).
\end{array}
\]
This program consists of two rules. The first rule serves as the base case, initializing the $\textit{TC}$ relation with all explicitly defined edges. The second rule is recursive and is applied iteratively to the currently derived facts, monotonically accumulating new tuples until the system reaches a global fixpoint at which no further facts can be generated.

Real-world Datalog programs, such as those in the \texttt{doop} analysis suite, consist of hundreds of rules with complex inter-dependencies. To schedule these systematically, the compiler constructs a dependency graph of all relations. Rules that are mutually recursive (\emph{i}.\emph{e}., forming cyclic dependencies) are grouped into a strongly connected component (SCC). The program is first stratified, and the engine then evaluates it stratum by stratum, which is known as \emph{stratification}~\cite{abiteboul1995foundations}. Each SCC is evaluated iteratively until it reaches its local fixpoint, at which point its output relations are finalized and subsequently treated as static, read-only inputs for all downstream strata.

The standard approach to implementing a Datalog engine is to compile logical rules into positive relational algebra operations ($\mathcal{RA}^+$)~\cite{ceri1986translation}. Variables shared across multiple body clauses implicitly act as equi-join constraints. For instance, the recursive rule in the TC example translates into a relational join followed by a union. The iterative semantics are captured by an additional fixpoint operator $\mathcal{O}$ : $\textit{TC}_{next} = \mathcal{O}(\textit{TC} \cup (\pi_{x,y}(\textit{TC} \bowtie \textit{Edge})))$. 
However, re-evaluating the full $\textit{TC}$ relation on every iteration redundantly re-discovers previously known paths. To achieve tractable performance, modern engines exclusively use \emph{semi-na\"ive evaluation}~\cite{ceri1989you}, rewriting the fixpoint computation into a differential form. By isolating the facts ($\Delta$) derived in previous iteration from the accumulated full facts ($\textit{TC}_{full}$), a rule only fires if at least one recursive body clause binds to a novel tuple from the immediate prior iteration. The execution is thus transformed into:
\[
\Delta \textit{TC}_{i+1} = \pi_{x,y}(\Delta \textit{TC}_i \bowtie \textit{Edge}) \setminus \textit{TC}_{full}
\]
This differential equation establishes the core execution pipeline of Datalog: compute the join over the delta, deduplicate the results via set difference, and structurally \textit{merge} the novel tuples back into the full relation for the next iteration.

Physically, the semi-na\"ive evaluation loop alternates between intensive read and write operations. Join processing is dominated by index probes and memory reads, whereas delta generation and index updates at the end of each iteration are fundamentally write-intensive. GPUs have emerged as natural accelerators for these workloads, since their high global memory bandwidth and thread latency hiding. Martinez et al.~\cite{martinez2013datalog} built early GPU Datalog prototypes, though they did not yet outperform the long-standing CPU gold standard, Souffl\'e~\cite{jordan2016souffle}. More recently, by tuning data structures and adopting hardware-aware, column-oriented memory layouts, GPU engines like GDlog~\cite{sun2025optimizing} and VFLog~\cite{sun2025column} have demonstrated speedups of more than 10$\times$ over CPU baselines. Lobster~\cite{lobster} extends this line of work to richer semantics, including provenance semirings, and targets neuro-symbolic learning workloads.



\paragraph*{Join Processing and Worst-Case Optimal Join}
\label{sec:bg-wcoj}
The join operator is one of the most important components in Datalog engine design, and the choice of how that join is evaluated largely determines whether recursive evaluation is asymptotically efficient or incurs unnecessary work. Existing GPU Datalog engines~\cite{shovon2025multi, sun2025optimizing, sun2025columnorienteddatalog}, like earlier CPU systems
~\cite{jordan2016souffle}, evaluate multi-relation rule bodies using chained binary joins. A rule with $k$ body atoms is compiled into $k-1$ pairwise joins, each consuming the intermediate result produced by the previous stage. This is the standard relational evaluation strategy, and it is what makes the column-store machinery and bulk kernel pipelines of GDlog and VFLog directly applicable. However, it also inherits a well-known asymptotic weakness on cyclic queries.
\[
\textit{Triangle}(x,y,z) \leftarrow R(x,y),\; S(y,z),\; T(z,x)
\]
Based on the Triangle query as shown above, a binary plan first computes $R \bowtie S$ on the shared variable $y$, materializing every two-hop path, and only then probes $T$ to discard paths whose closing edge does not exist. When $R$ and $S$ share a high-degree vertex, the intermediate balloons to $\Theta(|E|^2)$ even though the AGM bound~\cite{atserias2013size} guarantees the upper bound size of the final output of $O(|E|^{3/2})$.

Worst-Case Optimal Joins (WCOJ)~\cite{ngo2018worst} were introduced to address this asymptotic gap. Rather than processing one relation pair at a time, WCOJ fixes a global variable order $(x_1,\ldots,x_m)$ and proceeds attribute at a time. At level $i$, it intersects the projections of all relations constraining $x_i$ under the current partial binding $(x_1,\ldots,x_{i-1})$, and only recurses into $x_{i+1}$ along surviving values. For the triangle, the algorithm intersects $R.x \cap T.x$ first, prunes any $x$ that cannot participate in a triangle, then for each surviving $x$, we subset $R$ based on the survived $x$ and intersects $R[x].y \cap S.y$, and finally intersects $S[y].z \cap T[x].z$. Branches that cannot complete into a full output tuple are discarded before they are materialized, and the total work is provably bounded by the AGM bound on the output size.

One classic implementation of WCOJ is Leapfrog Triejoin (LFTJ)~\cite{veldhuizen2014leapfrog}. LFTJ requires relations to be stored in a sorted structure and performs each level as a search-based intersection: every relation mentioning the current variable exposes a sorted unary iterator with a seek operation, and the algorithm advances the iterators in lockstep by repeatedly seeking each iterator to at least the maximum key currently held by the others, until all iterators agree on a common key or one iterator is exhausted. Free Join~\cite{wang2023free} later unified WCOJ and binary joins within a single plan space. Crucially, LFTJ is defined over an \emph{interface}---sorted unary iteration with forward seek---not a physical layout. The original LogicBlox implementation uses B-trees, but any structure that supports this interface, including sorted arrays, can serve as the underlying storage. A parallel line of work, including Generic Join~\cite{ngo2018worst} and Free Join~\cite{wang2023free}, realizes the same attribute-at-a-time evaluation over hash tries. Free Join further unifies WCOJ and binary joins within a single framework.




\section{Challenge: Parallel WCOJ for Datalog}
\label{sec:wcoj-on-gpu-datalog}

While Worst-Case Optimal Joins theoretically resolve the asymptotic memory blowups of binary joins (\S\ref{sec:bg-wcoj}), translating this attribute-at-a-time execution model to SIMT hardware introduces a fundamental system design tension. 
Existing systems---spanning modern recursive GPU Datalog and static GPU WCOJ implementation---each compromise on a fundamental requirement of efficient Datalog evaluation. Consequently, they remain structurally incapable of executing complex, multi-way recursive queries end-to-end, failing on the demanding real-world applications.

\paragraph{The VRAM Wall of Binary Joins.} \label{sec:vram-wall}
Existing recursive GPU Datalog systems (e.g., GDlog~\cite{sun2025optimizing}, VFLog~\cite{sun2025column}) rely exclusively on chained binary-join evaluation. To maintain balanced SIMT thread scheduling, they fully materialize intermediate relations in device memory at every body clause, trading memory capacity for skew-tolerance and execution determinism. While effective for simple graph queries (e.g., Transitive Closure), this physical plan mathematically collapses on static program analysis workloads. These domains feature hundreds of cyclic, 6-to-8-way join rules. On these topologies, binary evaluation guarantees asymptotic blowups in intermediate relation sizes regardless of the chosen join order. Materializing these intermediates is no longer a performance tradeoff but an impossibility, instantly exhausting the VRAM capacity of any modern GPU.

\paragraph{The WCOJ Load Balancing Crisis.} \label{sec:load-balancing-crisis}
One effective way to bound this intermediate size algorithmically is using WCOJ. However, mapping WCOJ to a GPU exposes a fundamental mechanical vulnerability: it is highly vulnerable to the extreme load skew generated by power-law degree distributions. The most straightforward parallelization strategy assigns one outermost join variable to each thread group. In power-law graphs, if an inner variable binds to a ubiquitous high-degree hub, the evaluating thread group is trapped processing a massive, localized intersection space. This induces severe macroscopic load imbalance, leaving the vast majority of the GPU's Streaming Multiprocessors (SMs) idle while a few struggle with disproportionate workloads. 

\paragraph{The Index Rebuilding Overhead of Static GPU WCOJ} \label{sec:index-rebuild-price}
Existing static parallel WCOJ implementations (\emph{e}.\emph{g}., HoneyComb~\cite{wu2025honeycomb}, cuMatch~\cite{park2025cumatch}) mitigate this skew by preprocessing domains into spatial grids or heavily compressed multi-level prefix trees (e.g., CoCo tries).
By structuring the search space hierarchically, these indices naturally fracture massive, skewed root keys into smaller sub-trees at deeper levels by grouping child edges into discrete lexicographical intervals, allowing the static load balancing to distribute a single heavy-hitter's fan-out across multiple thread groups.
Because these structures flatten hierarchical relationships into tightly packed offset arrays to maximize read bandwidth, they are inherently immutable. While highly effective for one-shot queries over static graphs, this immutability fundamentally clashes with iterative Datalog. Semi-na\"ive evaluation injects new delta tuples into the working set on every single fixpoint iteration. While a flat column store absorbs these deltas via a lightweight, single-pass merge, the tightly packed hierarchical indices required by static implementations must be continuously reconstructed. As demonstrated in our micro-benchmark, this rebuild phase consumes up to 93\% of the total merge time, transforming a theoretically lightweight update into a bottleneck across hundreds of iterations.

\paragraph{Dynamic WCOJ Under Deep Skew} \label{sec:fraigile-ws}
Conversely, dynamic WCOJ systems (e.g., ALFTJ~\cite{lai2022accelerating}) adapt to skew at runtime by pushing heavily skewed sub-ranges into a global, lock-free queue for idle threads to steal. While functional for shallow, 3-way queries (e.g., triangle counting), this reactive paradigm is structurally fragile on the deep, 6-to-8-way query topologies of complex real-world queries like program analysis. Work-stealing heuristics fundamentally rely on localized thresholds, but an apparently modest fan-out at an outer join level can unpredictably cascade into a massive exponential explosion deep within the inner relations. This creates an impossible tuning paradox: setting a fine-grained donation threshold floods the global queue, collapsing the GPU under catastrophic atomic contention, while a coarse threshold fails to capture deeply buried skew, leaving threads trapped in massive localized workloads that may themselves need to be rebalanced again. Furthermore, even if the compute could be balanced, dynamic execution violates a strict requirement of Datalog: final results of join must be physically materialized. Because work-stealing prevents thread groups from determining their output bounds \emph{a priori}, threads must rely on contested global atomic counters to dynamically allocate write offsets. This introduces severe SIMT serialization, destroying the coalesced memory bandwidth required to sustain recursive evaluation.

\paragraph*{Architectural Imperatives.}
Because all prior approaches relax a constraint that recursive program analysis cannot afford to lose, \tool{} is designed by refusing to compromise. A viable engine must satisfy \textbf{three architectural imperatives}: 
\textbf{(1)} it must employ a two-phase, count-and-materialize pipeline that produces deterministic write offsets without atomic serialization (\S~\ref{sec:system-overview}); \textbf{(2)} it must utilize flat, sorted-column storage that admits cheap iterative delta merges without auxiliary index rebuilds (\S~\ref{sec:system-overview}); and \textbf{(3)} it must utilize a launch-time load-balancer that resolves outermost skew without runtime coordination (\S~\ref{sec:gpu-wcoj}).


\section{\tool{}: WCOJ on the GPU} \label{sec:system-overview}

\begin{figure}
    \centering
    \includegraphics[width=\linewidth]{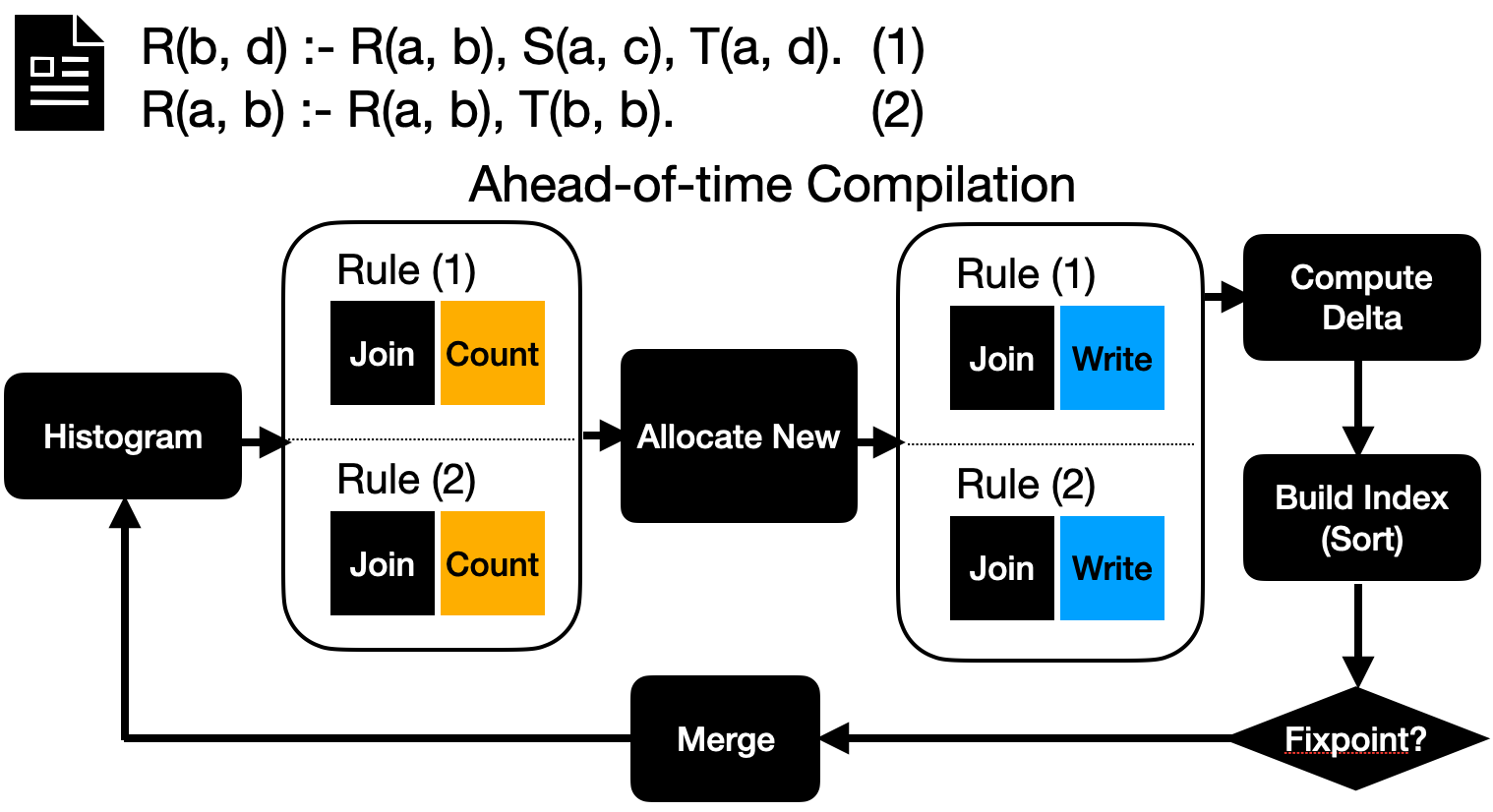}
    \caption{Architecture of a single \tool{} semi-na\"ive iteration. The AOT compiler lowers Datalog rules (top left) into batched \emph{Count} and \emph{Materialize} kernels.}
    \label{fig:architecture}
\end{figure}

\tool{} is an iterative, GPU-accelerated Datalog engine explicitly architected to satisfy the three execution imperatives established in \S\ref{sec:wcoj-on-gpu-datalog}. Rather than relying on heavy static index partitioning or dynamic work-stealing queues, \tool{} adheres strictly to semi-na\"ive evaluation semantics by mapping logical Datalog operations directly to physical hardware phases. As illustrated in Figure~\ref{fig:architecture}, an Ahead-of-Time (AOT) compiler translates high-level rules into a staged pipeline of CUDA kernels rather than interpreting queries at runtime. Within a single fixpoint iteration, this pipeline executes sequentially: it first employs a \textit{Histogram} phase to load-balance the outermost join (\S\ref{sec:gpu-wcoj}), followed by deterministic two-phase WCOJ rule evaluation (\textit{Count} and \textit{Materialize}). Finally, the engine isolates unique new derivations (\textit{Compute Delta}), enforces flat columnar layout requirements (\textit{Build Index}), and updates the persistent relational state (\textit{Merge}).


To bypass the SIMT serialization penalty of dynamic work-stealing queues (\textbf{Imperative 1}), \tool{} abandons runtime output allocation in favor of a strict two-phase join execution model. GPU architectures require lock-free memory writes to maximize global memory bandwidth; each thread must deterministically know its output offset before computation begins. Execution therefore begins with a \textit{Histogram} pass to bound the workload. Guided by these bounds, the \textit{Count Kernel} performs multi-way column intersections without materializing results, generating a thread-group-level histogram of output sizes. A global prefix sum over these counts dictates a bulk memory allocation for the iteration and computes deterministic write offsets for every thread. With memory pre-allocated and offsets synchronized, the \textit{Materialize Kernel} re-executes the intersection logic, writing the output tuples in a cooperative and coalesced manner.

To eliminate the crippling index rebuild tax of static WCOJ implementations (\textbf{Imperative 2}), \tool{} must support this two-phase pipeline without relying on immutable hierarchical structures (e.g., CoCo tries). Therefore, \tool{} represents all relations strictly as flat, sorted arrays utilizing a Structure-of-Arrays (SoA) layout. Each relation is sorted lexicographically in radix order according to the column access sequence dictated by the query plan. This ordering ensures that all tuples sharing a common prefix are stored contiguously in memory. By coupling this flat layout with a specialized join algorithm and a collaborative fast-skip mechanism, \tool{} emulates trie-like multi-way navigation directly over raw columns, completely bypassing the need to maintain auxiliary metadata.

The flat SoA layout is well-suited for the GPU computation and the semi-na\"ive execution loop. Following the \textit{Materialize Kernel}, \tool{} leverages this layout to execute the final stages of the pipeline (Figure~\ref{fig:architecture}). The materialized tuples enter the \textit{Compute Delta} phase, which filters the output against existing full version relations to isolate new tuples. To prepare the delta tuple for subsequent iterations, \tool{} subjects the new tuples to a sorting pass (\textit{Build Index}), a memory-bound operation suited for GPUs~\cite{satish2009designing}. Sorting organizes the newly generated tuples into the lexicographical SoA format required for contiguous range intersections in future joins. 
If the deduplicated delta is empty, execution terminates.
Otherwise, \tool{} proceeds to the \textit{Merge} phase. A naive physical implementation of this step would suffer from severe memory bandwidth waste during the rapidly decaying ``long tail'' of semi-na\"ive evaluation, where merging a microscopic delta would force a full structural rewrite of a massive base relation. To resolve this without sacrificing the simplicity of the WCOJ pipeline, \tool{} physically partitions each relation into a lightweight \emph{head} buffer and a massive fully-sorted \emph{body}. During the \textit{Merge} phase, incoming deltas are cheaply integrated strictly into the small head array. \tool{} only triggers a heavy, single-pass GPU path merge~\cite{green2012gpu} to flush the head into the body when a capacity threshold is reached. Because the WCOJ intersection iterators simply treat the head and body as a logical union during the join phases, this amortized design successfully absorbs long-tail write amplification while completely avoiding the index reconstruction tax of static WCOJ implementations.



\section{Work-Balanced WCOJ on the GPU}
\label{sec:gpu-wcoj}

To mitigate the load skew of power-law degree distributions without introducing too much runtime cross-thread group coordination (satisfying \textbf{Imperative 3}), \tool{} employs a \emph{root-level, histogram-guided WCOJ load balancing}. As demonstrated in \S\ref{sec:wcoj-on-gpu-datalog}, real Datalog workloads cannot be partitioned na\"ively along the outermost joined variable using a one-value-per-thread-group assignment. Rather than structurally reorganizing the relational data or relying on dynamic work-stealing queues, \tool{} relies on the \textit{Histogram} pass shown in Figure~\ref{fig:architecture} to compute a prefix-summed degree distribution of the outermost join column. To ensure this scheduling step remains genuinely lightweight across hundreds of fixpoint iterations, \tool{} does not recompute this histogram from scratch over the massive full relations. Instead, it maintains the histogram incrementally: the distribution is computed solely over the newly generated delta tuples, and then cheaply merged with the base relation's existing histogram alongside the data itself during the amortized \textit{Merge} phase. This incrementally maintained histogram flattens the highly skewed join space into a uniform 1-D array of work units, allowing the orchestrator to deterministically assign proportionally balanced slices of the root-level search space to thread blocks at kernel launch. The inner intersection levels then proceed as a standard, thread-cooperative Leapfrog Triejoin over the flat columns.

\begin{figure}
    \centering
    \includegraphics[width=1\linewidth]{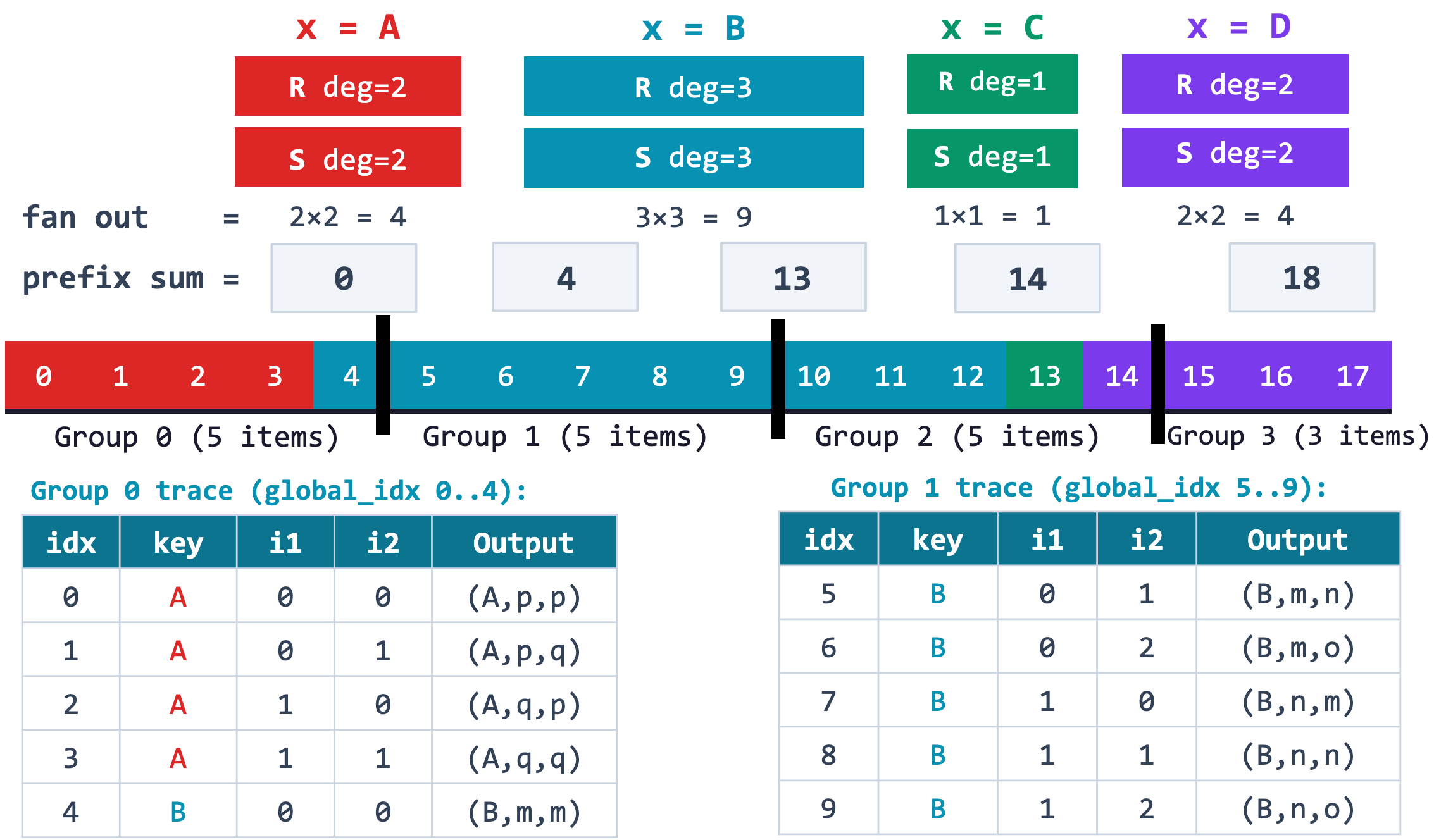}
    \caption{Histogram-guided partitioning flattens per-key fan-outs into a prefix-summed 1-D work space.}
    \label{fig:bg-rebalance}
\end{figure}

\begin{algorithm}[t]
\caption{Work-Balanced GPU WCOJ}\label{alg:bg-wcoj}
\begin{algorithmic}[1]
\Require Relations $R_1, \ldots, R_\ell$ as sorted column arrays;\;
variable order $(x_1, \ldots, x_m)$;\;
$p$ blocks, $w$ warps/block

\Statex \textbf{Phase 1: Histogram} (one warp per root key)
\State $U[0..K) \gets$ unique values of $x_1$ across sources
\ForAll{$i \in [0, K)$ \textbf{in parallel}}
\State $\mathcal{W}[i] \gets \textsc{Degree}(U[i])$
  \Comment{Fan-out under root key $i$}
\EndFor
\State $\mathcal{C} \gets \textsc{PrefixSum}(\mathcal{W})$;\;
$T \gets \mathcal{C}[K{-}1]$

\Statex \textbf{Phase 2: Count kernel}
(Alg.~\ref{alg:lb-wcoj} with emit $\to$ \texttt{count++})
\State $\mathit{tc}[\,...] \gets$ per-thread output counts
\State $\mathit{offsets} \gets \textsc{PrefixSum}(\mathit{tc})$;\;
allocate output to $\sum \mathit{tc}$

\Statex \textbf{Phase 3: Materialize kernel}
\end{algorithmic}
\end{algorithm}

\begin{algorithm}[t]
\caption{HG-WCOJ Kernel (per block $b$)}\label{alg:lb-wcoj}
\begin{algorithmic}[1]
\Require Prefix sums $\mathcal{C}[0..K)$, total $T$, root keys $U$
\Require Variable order $(x_1, \ldots, x_m)$,
sources $S_j$ for each variable $x_j$

\State $[b_s, b_e) \gets$ block $b$'s slice of $[0, T)$
\Comment{$\lceil T/p \rceil$}
\State $\kappa \gets \Call{BinarySearch}{\mathcal{C},\; b_s}$
\Comment{Starting root key}

\For{each key $\kappa$ overlapping $[b_s, b_e)$}
\State Narrow first-source handle to warp's share of key $\kappa$

\State \textsc{Prefix}$(x_1 {=} U[\kappa])$ on all sources
  \Comment{Cooperative}

\For{each $x_2$ in $\textsc{Intersect}(S[x_1{\to}x_2])$ across sources}
  \State \textsc{Prefix}$(x_2)$; narrow child handles
  \For{each $x_3$ in $\textsc{Intersect}(S[x_1{,}x_2{\to}x_3])$}
    \State $\cdots$ \Comment{Recurse through variable order}
    \State At leaf: \textbf{emit}$(x_1, \ldots, x_m)$
  \EndFor
\EndFor
\EndFor
\end{algorithmic}
\end{algorithm}

To illustrate this flattening mechanism, Figure~\ref{fig:bg-rebalance} traces \tool{}'s load balancing phase on the outermost joined column of a generic query $\textit{O}(k, x, y, \ldots) \leftarrow \textit{R}(k, x), \textit{S}(k, y), \dots$ with four distinct root values $A$, $B$, $C$, $D$, each generating a total inner search space (fan-out) of 4, 9, 1, and 4 work units respectively.
Before the actual join begin,
a \emph{histogram} kernel  sweeps the participating relation's outer-most column once and writes the per-key fan-out into a flat array indexed by outer-column rank (top row of the figure). For delta relations, this histogram is computed directly over the outermost column using Thrust primitives. For the full relation, it is maintained incrementally by merging the delta histogram with the current histogram of the full relation.

A \emph{prefix sum} over this array (second row) flattens the entire join's work into a 1-D space of total size $T = 18$, with each key $k$ owning the contiguous range $[\mathcal{C}[k{-}1], \mathcal{C}[k])$. In this example, there are four thread groups, so $p = 4$. The work-balanced \emph{count/materialize kernels} then assign each block-group a contiguous slice $\lceil T/p \rceil = 5$ of this flat space, regardless of how many distinct outer-column values fall inside the slice.
Group~0 receives work units $[0, 5)$, which spans key $A$ in full and the first work unit of key $B$; group~1 receives $[5, 10)$, entirely inside the heavy key $B$; group~2 covers the rest of $B$ together with $C$; group~3 covers $D$.
The heaviest key $B$, which under a naive one-key-per-group assignment would have monopolized a single block-group with nine sequential work units, is now split across three groups.

To recover the actual tuples to emit, each thread translates its global work-unit index back into a $(\mathit{key},\, i_1,\, i_2)$ triple by binary-searching the prefix array $\mathcal{C}$ for the owning key, then computing
$\mathit{local} = \mathit{global} - \mathcal{C}[\mathit{key}{-}1]$,
$i_1 = \mathit{local} \div d_2(\mathit{key})$,
$i_2 = \mathit{local} \bmod d_2(\mathit{key})$,
where $d_2(\mathit{key})$ is the inner-relation degree under that key (read from the per-key array).
The two indices $i_1$ and $i_2$ are then point lookups directly into the sorted columns of \textit{Outer} and \textit{Inner} --- no auxiliary index, no per-thread iterator state, no cross-warp coordination.
The bottom of Figure~\ref{fig:bg-rebalance} shows the resulting traces for groups~0 and~1: every flat work unit maps to exactly one output tuple, and adjacent threads within a warp emit adjacent positions of the inner relation, preserving the coalesced write pattern that the count/materialize pipeline (\S\ref{sec:system-overview}) requires.
The same decomposition extends to multi-way joins by treating the inner side as the cross-product of the remaining join levels under the chosen variable order.


For exceptionally deep join trees, single-level balancing is insufficient. For example, the recursive $\textit{CallGraphEdge}$ ($\textit{CGE}$) rule from micro-\texttt{doop} (Figure~\ref{fig:cge-split-compare}a) intersects seven logical variables across six body clauses; compiling this directly generates a six-level nested intersection loop. On a binary-join engine like VFLog, evaluating this requires decomposing it into five chained operators and materializing four intermediate relations, triggering memory blowouts because some clauses generate massive cross-products. While a monolithic WCOJ pipeline prevents this memory explosion, the six-level nesting exhausts the hardware register budget and exposes unmitigated fan-out skew buried in the inner logical layers.

To resolve this tension, \tool{} employs a targeted rewriting strategy known as \emph{helper-relation splitting} (Figure~\ref{fig:cge-split-compare}b). Rather than aggressively factoring the rule into a chain of shallow binary joins, \tool{} surgically isolates only the sub-trees responsible for catastrophic inner-level skew. In the original $\textit{CGE}$ rule, the type resolution variables ($\mathit{sn}$, $\mathit{dsc}$, $\mathit{t}$) frequently bind to ubiquitous base classes (e.g., \textsf{java.lang.Object}), generating severe unmitigated fan-out deep in the intersection pipeline. By factoring only these specific clauses into an independent $\textit{HelpNT}$ relation, the previously buried skew keys ($\mathit{sn}$, $\mathit{dsc}$, $\mathit{h}$) are exposed as top-level columns in the newly generated rule. This allows them to be aggressively balanced by \tool{}'s histogram-guided load balancing. Crucially, because the remaining clauses ($\textit{Reachable}$, $\textit{Instruction\_Method}$) do not induce catastrophic skew, they remain fused in the consumer rule. By strategically materializing a single boundary, \tool{} strikes a deliberate balance: it recovers the SIMT load-balancing efficacy lost to deep inner-level skew, while successfully evaluating the rest of the topology as a massive 5-way WCOJ kernel to preserve strict asymptotic memory efficiency.

\begin{figure}[t]
\centering

\textbf{(a) Original Monolithic Rule}
\begin{equation}\label{eq:cge-fused}
\begin{array}{l}
\textit{CallGraphEdge}(\mathit{i}, \mathit{m}) \leftarrow \\
\quad \textit{Reachable}(\mathit{j}), \textit{Instruction\_Method}(\mathit{i}, \mathit{j}),\\
\quad \textit{VirtualMethodInvoc}(\mathit{i}, \mathit{b}, \mathit{sn}, \mathit{dsc}),\\
\quad \textit{VarPointsTo}(\mathit{h}, \mathit{b}), \textit{HeapAllocation\_Type}(\mathit{h}, \mathit{t}),\\
\quad \textit{MethodLookup}(\mathit{sn}, \mathit{dsc}, \mathit{t}, \mathit{m}).
\end{array}
\end{equation}

\hrule
\vspace{1ex}

\textbf{(b) Targeted Structural Split}

\begin{equation}\label{eq:cge-helper}
\begin{array}{l}
\textit{HelpNT}(\mathit{sn}, \mathit{dsc}, \mathit{m}, \mathit{h}) \leftarrow \\
\quad \textit{MethodLookup}(\mathit{sn}, \mathit{dsc}, \mathit{t}, \mathit{m}),\\
\quad \textit{HeapAllocation\_Type}(\mathit{h}, \mathit{t}).
\end{array}
\end{equation}


\begin{equation}\label{eq:cge-consumer}
\begin{array}{l}
\textit{CallGraphEdge}(\mathit{i}, \mathit{m}) \leftarrow \\
\quad \textit{Reachable}(\mathit{j}), \textit{Instruction\_Method}(\mathit{i}, \mathit{j}), \\
\quad \textit{VirtualMethodInvoc}(\mathit{i}, \mathit{b}, \mathit{sn}, \mathit{dsc}),\\
\quad \textit{VarPointsTo}(\mathit{h}, \mathit{b}), \textit{HelpNT}(\mathit{sn}, \mathit{dsc}, \mathit{m}, \mathit{h}).
\end{array}
\end{equation}

\caption{$\textit{CallGraphEdge}$ rule from micro-DOOP before~(a) and after~(b) \tool{}'s targeted split.}
\label{fig:cge-split-compare}
\end{figure}

\section{Stream-Order Rule Parallelism}
\label{sec:stream-order}

Datalog is inherently monotonic. According to the CALM theorem~\cite{hellerstein2020keeping}, such monotonic computations guarantee deterministic convergence regardless of execution order, effectively eliminating the need for strict sequential coordination across different Datalog rules within the same stratum. \tool{} exploits this theoretical foundation to extract rule-level parallelism. Rather than evaluating rules sequentially, the \tool{} Ahead-of-Time (AOT) compiler orchestrates a phase-aligned schedule that interleaves execution at the granularity of individual pipeline phases (Count $\to$ Scan $\to$ Resize $\to$ Materialize) across independent CUDA streams. As illustrated in the center of Figure~\ref{fig:architecture}, the \textit{Count} and \textit{Materialize} kernels for independent rules (e.g., Rule 1 and Rule 2) are grouped by phase and dispatched concurrently onto separate CUDA streams, allowing the hardware scheduler to seamlessly overlap their execution and new tuple memory allocation on the device.

This stream-ordered model directly resolves the interacting bottlenecks of hardware underutilization inherent to real-world workloads like the DOOP program analysis suite. Individual rules frequently lack sufficient workload volume to fully saturate device Streaming Multiprocessors (SMs), particularly during the rapidly decaying ``delta tail'' of later fixpoint iterations. Executing these small kernels sequentially leaves hardware idle and exposes the pipeline to severe CUDA kernel launch and device memory allocation overheads. Furthermore, complex Datalog rules may contain up to eight joined body clauses. As shown in \S~\ref{sec:bg-datalog}, each clause compiles into a nested intersection loop, dictating that intermediate materialized values and probing iterators reside in fast on-chip registers. This deep nesting generates intense register pressure, which limits the number of active warps per SM and prevents the hardware scheduler from hiding global memory latency. Stream ordering explicitly resolves this trap: by dispatching independent rules concurrently across streams, \tool{} amortizes launch and allocation overheads, saturates the SMs with independent active warps, and sustains high device occupancy.

While CPU-based incremental Datalog systems, such as FlowLog~\cite{zhao2025flowlog}, leverage CALM further and achieve fully asynchronous, cross-iteration parallelism, this tuple-at-a-time streaming model is fundamentally incompatible with GPU architectures. SIMT throughput relies strictly on dense bulk operations, such as single-pass radix sorts, two-buffer path merges, and coalesced count/materialize kernels, to saturate global memory bandwidth. Fully asynchronous execution generates scattered, uncoalesced memory writes that destroy this throughput. By restricting concurrency to phase-aligned, intra-iteration execution, \tool{} extracts the hardware-friendly parallelism permitted by monotonicity while strictly preserving the dense, bandwidth-optimized data paths required for GPU acceleration.

\begin{table}[t]
\centering
\caption{Feature comparison of evaluated systems.}
\label{tab:feature-comparison}
\small
\resizebox{\columnwidth}{!}{%
\begin{tabular}{lcccccc}
\toprule
 & \textbf{\tool{}} & \textbf{VFLog} & \textbf{Souffle} & \textbf{Ascent} & \textbf{FlowLog} & \textbf{cuMatch} \\
\midrule
GPU          & \cmark & \cmark & \xmark & \xmark & \xmark & \cmark \\
WCOJ         & \cmark & \xmark & \xmark & \xmark & \xmark & \cmark \\
Semi-na\"ive & \cmark & \cmark & \cmark & \cmark & \cmark & \xmark \\
Recursion    & \cmark & \cmark & \cmark & \cmark & \cmark & \xmark \\
Negation     & \cmark & \xmark & \cmark & \cmark & \cmark & $\triangle$ \\
Macro-based  & \cmark & \xmark & \xmark & \cmark & \xmark & \xmark \\
Incremental  & \xmark & \xmark & \xmark & \xmark & \cmark & \xmark \\
\bottomrule
\end{tabular}%
}

{\footnotesize \cmark\ supported, \xmark\ not supported, $\triangle$ partial.}
\end{table}


\begin{table*}[t]
\centering
\caption{Execution time (seconds) across different workloads. \tool{} runs on a single NVIDIA RTX 6000 Ada GPU. CPU engines run on a 32-core AMD EPYC 9655 CPU with 256\,GB RAM. Souffl\'e times exclude CSV data-loading overhead, measured separately via load-only programs. \textbf{SF} abbreviate Souffl\'e, \textbf{Asc} abbreviate Ascent, \textbf{FL} abbreviate Flowlog.}
\label{tab:benchmark-results}
\begin{tabular}{ll rr rrrr rrr}
\toprule
\textbf{Workload} & \textbf{Dataset} & \textbf{Input} & \textbf{Iters} & \textbf{\tool{}} & \textbf{Ascent} & \textbf{FlowLog} & \textbf{Souffl\'e} & \textbf{vs Asc} & \textbf{vs FL} & \textbf{vs SF} \\
                   &                  &                &                & \textbf{(1 GPU)}   & \textbf{(32c)} & \textbf{(32c)}   & \textbf{(32c)}     &                 &                &                \\
\midrule
\multirow{4}{*}{ddisasm}   & cvc5       &   1.5M &  751 &  \textbf{0.73} &   7.5  &  11.8  &   7.7  &  10$\times$ &  16$\times$ &  11$\times$ \\
                           & z3         & 126.9M & 1542 &  \textbf{2.00} &  24.4  &  14.7  &  10.7  &  12$\times$ & 7.4$\times$ & 5.4$\times$ \\
                           & coqidetop  &  31.6M & 1192 &  \textbf{1.07} &   4.9  &   2.9  &   6.4  &   5$\times$ & 2.7$\times$ & 6.0$\times$ \\
                           & lean4      &  32.8M &  530 &  \textbf{0.61} &   5.3  &   4.3  &  13.9  &   9$\times$ & 7.0$\times$ &  23$\times$ \\

\midrule
\multirow{4}{*}{doop}      & batik     &  20.4M &  248 &  \textbf{1.93} &  22.8  &  11.9  &  22.2  &  12$\times$ &   6$\times$ &  12$\times$ \\
                           & eclipse   &   3.3M &  253 &  \textbf{1.70} &  16.2  &  12.4  &  15.2  &  10$\times$ &   7$\times$ & 8.9$\times$ \\
                           & biojava   &  11.1M &  179 &  \textbf{1.60} &  40.6  &   8.3  &  45.0  &  25$\times$ & 5.2$\times$ &  28$\times$ \\
                           & zxing     &   5.5M &  167 &  \textbf{0.89} &  11.4  &   8.5  &  31.4  &  13$\times$ &  10$\times$ &  35$\times$ \\
\midrule
\multirow{4}{*}{polonius}  & clap-rs   &   1.1M & 1488 &  \textbf{2.64} &  70.9  &  33.5  & 135.9  &  27$\times$ &  13$\times$ &  51$\times$ \\
                           & wgpu      &   1.1M & 1488 &  \textbf{2.70} &  69.5  &  27.8  & 119.2  &  26$\times$ &  10$\times$ &  44$\times$ \\
                           & materialize &  17.8M & 2322 &  \textbf{2.29} &  26.5  &  99.1  &  41.9  &  12$\times$   &  43$\times$ &  18$\times$ \\
                           & scallop     &  28.4M & 1360 &  \textbf{0.88} &  18.1  &  38.4  &   8.4  &  21$\times$   &  44$\times$ & 9.5$\times$ \\
\midrule
\multirow{2}{*}{andersen}  & medium    &  34.0M &   25 &  \textbf{0.06}&   2.0  &   2.7  &  10.6  &  32$\times$ &  43$\times$ & 176$\times$ \\
                           & large     &  68.0M &   24 &  \textbf{0.12} &   4.2  &   5.1  &  25.7  &  35$\times$ &  43$\times$ & 214$\times$ \\
\midrule
\multirow{2}{*}{CSPA}      & httpd     &   1.5M &   31 &  \textbf{0.73} &  34.7  &  13.9  &  31.3  &  48$\times$ &  19$\times$ &  43$\times$ \\
                           & postgresql&   4.7M &   34 &  \textbf{0.96} &  36.1  &  14.1  &  38.0  &  38$\times$ &  15$\times$ &  40$\times$ \\
\midrule
galen                      & galen     & 977K &   33 &  \textbf{0.28} &   7.0  &   6.3  &  20.6  &  25$\times$ &  23$\times$ &  74$\times$ \\
\bottomrule
\end{tabular}
\end{table*}

\section{Evaluation}


We evaluate \tool{} across two distinct hardware settings, establishing a cost-equivalent baseline to isolate the architectural difference of GPU-accelerated evaluation against state-of-the-art CPU executions. GPU experiments are conducted on a workstation equipped with a single NVIDIA RTX 6000 Ada Generation (48\,GB) paired with an AMD Ryzen Threadripper PRO 5945WX (12 physical cores). CPU baselines are evaluated on a cloud instance with an AMD EPYC 9655, utilizing 32 exposed physical cores and 256\,GB of DDR5 memory. These two platforms are strictly matched by commodity cloud rental cost ($\approx$\$1/hour at the time of writing). While they are not directly comparable on raw power budget---the entire EPYC draws roughly 400\,W  while we only own partial of it, against the RTX 6000 Ada's 300\,W---their power consumption are at same grade.

As summarized in Table~\ref{tab:feature-comparison}, we evaluate \tool{} against five external systems that collectively span the architectural design space of modern relational and graph query evaluation. For GPU-accelerated baselines, we compare against cuMatch~\cite{park2025cumatch}, a WCOJ engine that accelerates static graph queries but lacks semi-na\"ive recursion, and VFLog~\cite{sun2025column}, a recursive Datalog engine structurally constrained by binary joins and materialized intermediates. Both GPU baselines, alongside \tool{}, are compiled with Clang 20 and the NVIDIA HPC SDK 25.5. For CPU baselines, we evaluate against FlowLog~\cite{zhao2025flowlog}, an incremental Datalog engine built atop Differential Dataflow; Souffl\'e~\cite{jordan2016souffle}, the de-facto standard ahead-of-time Datalog compiler; and Ascent~\cite{sahebolamri2022seamless}, a high-performance macro-based Rust engine. As Table~\ref{tab:feature-comparison} illustrates, \tool{} is uniquely positioned within this landscape as the only system uniting GPU acceleration, WCOJ execution, and full recursive Datalog semantics. All CPU-based baseline systems are built in their maximally optimized release configurations, with parallel evaluation strictly enforced at the highest hardware thread count exposed by the host platform.

\subsection{End-to-end comparison}
\label{sec:eval-end-to-end}

We evaluate the central architectural benefits of this paper: by structurally bypassing the intermediate memory exhaustion of binary joins, \tool{} enables high-performance, end-to-end execution of deep, multi-relation program analysis on a GPU. Prior GPU Datalog engines (e.g., GDlog~\cite{sun2025optimizing}, VFLog~\cite{sun2025column}) are excluded from Table~\ref{tab:benchmark-results} because their binary-join execution models reliably Out-Of-Memory (OOM) on these topologies, as mathematically established in \S\ref{sec:wcoj-on-gpu-datalog}. Consequently, we compare \tool{} against three mature CPU engines: Souffl\'e~\cite{scholz2016fast}, FlowLog~\cite{zhao2025flowlog} (built on Differential Dataflow~\cite{mcsherry2013differential}), and Ascent~\cite{sahebolamri2022seamless}. All baselines execute at full multi-core parallelism on cost-equivalent hardware. To isolate the core relational bottleneck, we evaluate the recursive structures of these workloads based on FlowLog's benchmark suite~\cite{zhao2025flowlog}, leaving peripheral language features (e.g., arithmetic, user-defined functors) to future engineering.

All reported measurements strictly evaluate the core compute phase, explicitly excluding initial data loading. For Ascent, timing begins exactly when the compiled \emph{run()} routine is invoked. For FlowLog, we report the system's internally profiled reasoning time. Because Souffl\'e's compiled binaries do not natively expose isolated execution timers, we derive its compute phase by measuring the total GNU \texttt{time} wall-clock duration and mathematically subtracting the execution time of an I/O-only program stripped of its recursive rules. 

Table~\ref{tab:benchmark-results} reports these isolated end-to-end execution times across seven workload classes and 17 datasets. \tool{} outperforms all baselines across every benchmark, demonstrating geometric mean speedups of 21$\times$ over Ascent, 14$\times$ over FlowLog, and 26$\times$ over Souffl\'e. Because this is a CPU-vs-GPU comparison, these massive gains reflect the combined synergy of \tool{}'s architectural mechanisms: raw GPU memory bandwidth, the columnar WCOJ execution model, flat-array delta merges, stream-order rule multiplexing, and histogram-guided skew mitigation. 

A deeper analysis of the speedup variance across workloads. On complex, multi-way program analysis workloads (\texttt{doop}, \texttt{ddisasm}, \texttt{polonius}), \tool{} achieves strong 3$\times$--51$\times$ speedups. The benchmark suite inherited from FlowLog specifically targets the critical relational bottlenecks within these industrial frameworks, evaluating foundational rules from micro-\texttt{doop} alongside subsets from the most computationally expensive Strongly Connected Components of \texttt{ddisasm}. These recursive topologies are fundamentally characterized by long-tail delta distributions: across their extensive fixpoint chains, the vast majority of iterations generate only a microscopic trickle of new delta tuples. This severe structural sparsity inherently bounds peak GPU efficiency, exposing the sequential launch overhead of the semi-na\"ive evaluation loop. However, while this specific topology allows CPU baselines like FlowLog to perform at their peak by amortizing iteration overhead via LSM-like versioning, \tool{}'s WCOJ microarchitecture fundamentally overpowers them, maintaining a decisive victory even in their strongest operational regime. Furthermore, as detailed in \S\ref{sec:eval-stream}, \tool{} successfully deploys stream-parallel multiplexing to mitigate this long-tail underutilization on highly fractured workloads like \texttt{doop}.

\texttt{Andersen} and \texttt{CSPA} yield \tool{}'s most overwhelming margins (32$\times$--214$\times$). Structurally, these workloads resemble non-linear Transitive Closure topologies~\cite{smart-tc}; they converge in very shallow fixpoint chains (24--34 iterations) but compute incredibly massive, dense relational intersections at every single step. This dense iterative behavior perfectly aligns with SIMT architectural strengths. Unencumbered by the sequential latency of long-tail fixpoints, \tool{} can fully saturate its thousands of SMs with WCOJ math, exposing the true hardware delta between GPU memory bandwidth and traditional CPU evaluation.

\subsection{\tool{} versus SOTA GPU-based systems}
As established in \S\ref{sec:wcoj-on-gpu-datalog}, existing GPU systems cannot execute our primary end-to-end program analysis workloads: static WCOJ engines lack recursive semi-na\"ive evaluation, and binary-join Datalog engines exhaust memory on complex multi-way rules. To rigorously evaluate \tool{}'s microarchitectural efficiency against these state-of-the-art (SOTA) baselines, we design two targeted experiments constrained strictly to their respective operational limits.

\paragraph{Versus GPU-based WCOJs}
First, we compare \tool{} against the fastest available GPU WCOJ system, cuMatch~\cite{park2025cumatch}. We evaluate both systems' join time (partition time required by cuMatch is not included in timing) on the three most time-consuming graph pattern matching queries (Q6, Q7, Q9) with representative query pattern (triangle with filter, star shape join, triangle with negation) from the LSQB benchmark~\cite{lsqb}. We exclude older systems like ALFTJ~\cite{lai2022accelerating}, as cuMatch has already demonstrated $>3$x speedups over them. To isolate raw in-memory intersection speed, we tune the dataset to Scale Factor 100 (SF=100), guaranteeing all computation remains strictly within GPU VRAM. Although cuMatch is explicitly architected to support out-of-core disk spilling---a design choice that inherently trades peak in-memory throughput for capacity---it serves as the most rigorous baseline available for evaluating multi-way GPU intersections.

\begin{figure}
    \centering
    \includegraphics[width=0.9\linewidth]{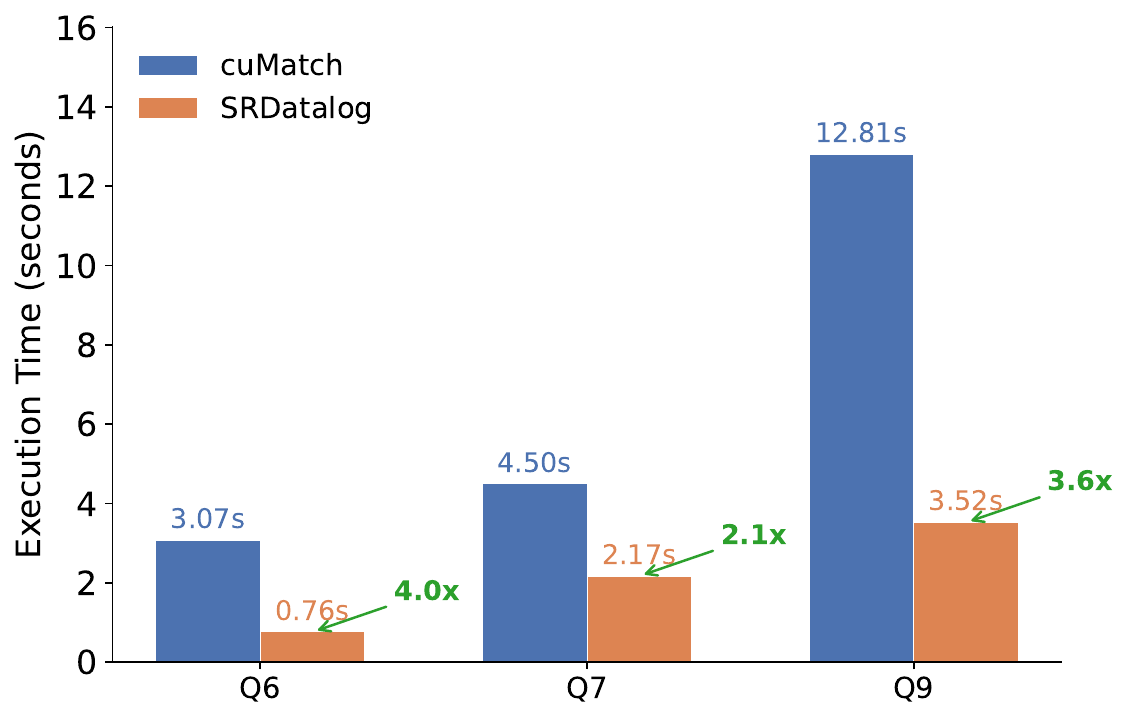}
    \caption{LSQB SF=100 execution time. \tool{} is 2.1--4.0$\times$ faster than cuMatch on all query types.}
    \label{fig:lsqb}
\end{figure}

As shown in Figure~\ref{fig:lsqb}, \tool{} achieves speedups of 2.1x to 4.0x across all three query topologies. The performance gap is most pronounced on Q9 (2-hop path with negation), where \tool{} reduces execution time from 3.07s to 0.78s (4.0x), and Q6 (2-hop path with interest), dropping execution from 12.85s to 3.64s (3.5x). These results validate \tool{}'s foundational storage design: by operating directly on flat columnar arrays, \tool{} avoids the massive pointer-chasing and allocation overheads required to traverse and maintain cuMatch's specialized grid-partitioned structures.

\paragraph{Versus GPU-Native Datalog Engines}
Because the recently proposed Lobster~\cite{lobster} engine remains closed-source, we isolate recursive overhead against VFLog~\cite{sun2025column}, the fastest open-source GPU Datalog alternative. Since VFLog reliably triggers out-of-memory (OOM) failures on deep program analysis rules, we evaluate both systems on the Same Generation (SG) benchmark. Featuring three body clauses, SG represents the precise topological threshold where binary evaluation is forced to physically materialize one intermediate relation per iteration. 
Table~\ref{tab:sg-benchmark} reports execution times across four real-world graphs. \tool{} outperforms VFLog by 2.3$\times$ to 7.1$\times$, converging on identical semi-na\"ive fixpoints. On the most iteration-heavy dataset, \texttt{vsp\_finan} (513 iterations), \tool{} reduces execution time from 35.5s to 5.01s. This performance delta directly exposes the severe memory bandwidth tax of binary joins: at every iteration, VFLog must physically write and subsequently read an intermediate relation to global VRAM. By employing WCOJ, \tool{} mathematically bypasses intermediate materialization, converting these global bandwidth savings directly into iteration speed. Furthermore, \tool{} matches and exceeds the 2$\times$--3$\times$ speedup over VFLog reported by Lobster~\cite{lobster}. \tool{} achieves these margins purely through algorithmic WCOJ efficiency, avoiding the severe memory capacity penalties associated with Lobster's mutable hashmaps.


\begin{table}[t]
\centering
\caption{Same Generation (SG) benchmark comparing \tool{} (abbreviated as \textbf{SR}) against VFLog (abbreviated as \textbf{VF}).
All experiments run on an NVIDIA RTX 6000 Ada.}
\label{tab:sg-benchmark}
\small
\setlength{\tabcolsep}{4.5pt}
\begin{tabular}{lrrrrrc}
\toprule
\textbf{Dataset} & \textbf{Edges} & $|\texttt{SG}|$ & \textbf{Iters} & \textbf{SR(s)} & \textbf{VF(s)} & \textbf{Speedup} \\
\midrule
fe-sphere      &   49{,}152 &  206M &  127 & \textbf{1.04} &  3.73 & 3.6$\times$ \\
SF.cedge       &  223{,}001 &  382M &  269 & \textbf{1.10} &  7.63 & 6.9$\times$ \\
fe\_body       &  163{,}734 &  408M &  125 & \textbf{3.23} &  7.56 & 2.3$\times$ \\
vsp\_finan     &  552{,}020 &  865M &  513 & \textbf{5.01} & 35.5  & 7.1$\times$ \\
\bottomrule
\end{tabular}
\end{table}


\subsection{Ablation Study: Handling Data Skew}

\begin{figure}[t]
\centering
\includegraphics[width=0.9\linewidth]{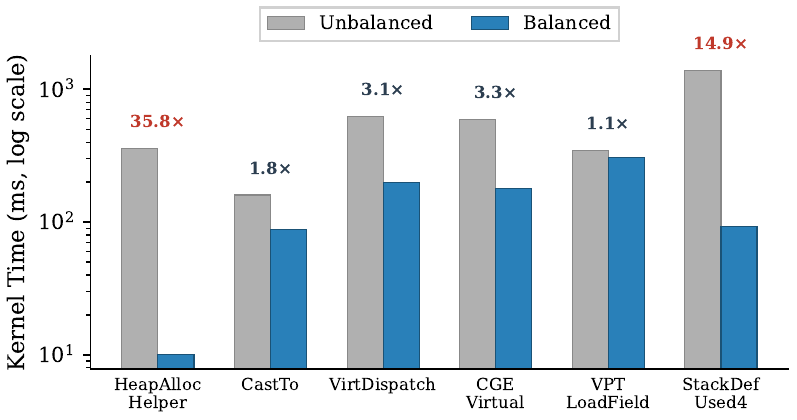}
\caption{Kernel microseconds time with (blue) and without (gray) skew handling across six rules (log scale). Kernel time measured by NVIDIA NSight System profiler.}
\label{fig:skew-handling}
\end{figure}

In Section~\ref{sec:gpu-wcoj}, we introduced \tool{}'s histogram-guided work partitioning. To isolate its benefits, we select several kernels from the micro-\texttt{doop} and compare performance with and without histogram-guided partitioning. For the baseline, we implement a code generation path in compiler that join kernels that assign naively one outermost unique value per thread group, and compare their kernel times against the balanced kernels emitted by \tool{}'s compiler. Kernel times are measured with the NVIDIA Nsight Systems profiler, and the results are reported in Figure~\ref{fig:skew-handling}.

Speedups range from $1.1\times$ to $35.8\times$, heavily dependent on whether data skew aligns with the outermost partitioned variable. \emph{HeapAllocHelper} achieves massive gains ($35.8\times$) by applying the structural splitting insight detailed in \S~\ref{sec:gpu-wcoj}. In real-world Java applications, a trivial number of base types (e.g., \textit{Object}, \textit{Exception}) absolutely dominate the subtyping hierarchy, creating severe inner-variable skew deep within the heap allocation control flow. By splitting this rule, \tool{} elevates these ubiquitous types to the root level, allowing the 1-D histogram to perfectly absorb the imbalance. In contrast, the two virtual-dispatch rules ($3.1\times$--$3.3\times$) exhibit moderate outer skew on the receiver pointer but suffer from unmitigated inner skew on the \textit{VarPointsTo} fan-out; thus, the speedup reflects only the resolution of the outermost imbalance. \emph{CastTo} ($1.8\times$) exhibits extreme outer skew---a single key encapsulates 75.5\% of the workload---but operates on a modest absolute data volume (35.8M tuples, 160\,ms baseline), fundamentally restricting the theoretical speedup ceiling. Finally, \emph{VPT\_LoadField} ($1.1\times$) highlights a deliberate scheduling trade-off. To minimize per-iteration overhead across hundreds of semi-na\"ive fixpoint steps, \tool{} strictly places the delta relation ($\Delta\textit{VarPointsTo}$) outermost. Because this specific delta is small and near-uniformly distributed, the root histogram provides neutral scheduling benefits. The actual skew resides within the inner \textit{InstanceFieldPointsTo} relation, demonstrating that while the delta-first policy is globally optimal for iteration efficiency, it leaves un-split inner skew deliberately unaddressed.

\subsection{Stream Parallel Kernel Scheduling Ablations}
\label{sec:eval-stream}

\begin{figure}
    \centering
    \includegraphics[width=0.9\linewidth]{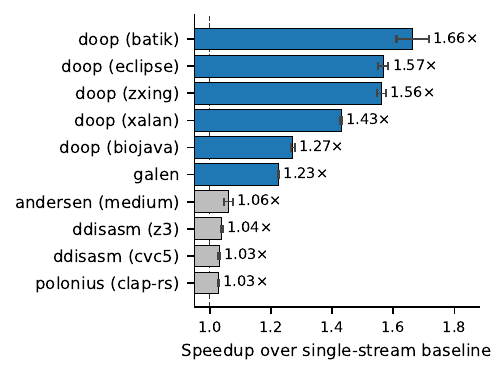}
    \caption{Wall-time speedup of the stream-parallel schedule over a sequential rule-by-rule schedule. Each bar is the ratio of the two means over five runs}
    \label{fig:ablation-stream}
\end{figure}

To isolate the performance impact of the phase-aligned schedule detailed in \S~\ref{sec:stream-order}, we compile each benchmark in two configurations: the proposed \emph{stream-parallel} schedule and a baseline \emph{sequential} schedule that emits a strict rule-by-rule evaluation loop onto the default CUDA stream. Figure~\ref{fig:ablation-stream} reports the wall-clock speedup of the stream-parallel approach.

Workloads characterized by stratum of large numbers of complex rules exhibit substantial performance gains. The five DOOP datasets achieve speedups ranging from $1.27\times$ (\emph{biojava}) to $1.66\times$ (\emph{batik}), while \emph{galen} accelerates by $1.23\times$. The dominant recursive stratum in micro-\texttt{doop} contains twelve independent rules handling distinct pointer aliasing cases (e.g., array accesses, static loads, local variable assignments). Each rule performs deep three-to-six-way intersections, inducing heavy register pressure, yet processes insufficient data volume to independently saturate the GPU. Hardware profiling via NVIDIA Nsight Systems confirms that stream multiplexing natively resolves this underutilization. On the \emph{batik} workload, the stream-parallel configuration accumulates 3.0\,s of raw device execution time but completes in only 2.3\,s of wall-clock time, achieving a sustained $1.30\times$ hardware concurrency factor by actively overlapping independent compute and memory allocation stalls.

The remaining four benchmarks (\emph{andersen}, both \emph{ddisasm} variants, and \emph{polonius}) exhibit negligible variance ($\pm 3\%$). This result directly reflects constrained workload where dominant strata contain three or fewer rules. (We note that while full \emph{ddisasm} comprises over 3,000 rules and structurally similar to \texttt{doop}, we evaluate the reduced artifact from FlowLog~\cite{zhao2025flowlog}, whose dominant strongly connected component happens to be dense but rule-sparse). In such tight configurations, overlapping trivial secondary rules fails to inject meaningful hardware concurrency. This limitation is most pronounced in \emph{andersen}, where a single massively recursive rule monopolizes execution. Because this isolated rule provides sufficient data volume to fully saturate device SMs on its own, no idle resources remain for concurrent streams to exploit. Ultimately, these results empirically validate our core thesis: stream multiplexing  recovers occupancy for broad, fractured strata (\emph{e}.\emph{g}., micro-\texttt{doop}), but yields diminishing returns for rule-sparse workloads that already saturate the hardware.

\section{Related Work}

\paragraph{High-performance Datalog systems} have been studied across both CPU and GPU platforms.
On the CPU side, systems such as Souffl\'e~\cite{jordan2016souffle} established
Datalog as a practical foundation for high-performance static analysis, while
RDFox~\cite{nenov2015rdfox} demonstrated that highly scalable rule processing can also be achieved in
RDF-oriented settings . On accelerators, early work
explored GPU-based Datalog execution~\cite{martinez2013datalog}, and more
recent systems have substantially improved GPU performance through
column-oriented designs~\cite{sun2025columnorienteddatalog}, optimized GPU
execution strategies~\cite{sun2025optimizing}, and scaling across multiple GPUs
and nodes~\cite{shovon2025multi}. Other recent frameworks have extended
GPU-accelerated Datalog-style reasoning to richer settings such as
neurosymbolic programming~\cite{lobster}, while FlowLog focuses on efficient
and extensible Datalog execution through incrementality~\cite{zhao2025flowlog}. These systems have advanced high-performance Datalog along several important axes, including scalable CPU execution, GPU acceleration, incrementality, and multi-device scaling. In contrast, our work focuses specifically on recursive semi-na\"ive Datalog with worst-case optimal
multi-way joins, where GPU execution must be tightly integrated with iterative
delta maintenance, deterministic materialization, and fixpoint evaluation.

\paragraph{Graph Analytics Engines}  Broader work on graph analytics and irregular GPU execution has explored several related directions. Prior systems have improved work efficiency for irregular vertex-centric graph processing on GPUs~\cite{graph_gpu_vora}; supported declarative and incremental analytics over evolving graphs~\cite{graphbolt,dzig,grafs}; and accelerated pattern-aware graph mining, subgraph query processing, and subgraph counting~\cite{peregrine,profilling_graph_martin,frienge_sgc}. These works highlight the importance of handling skew, irregular parallelism, and graph-structured computation efficiently on modern hardware. However, they generally target graph analytics or subgraph-oriented tasks rather than recursive semi-na\"ive Datalog evaluation. In contrast, our work focuses on recursive Datalog with worst-case optimal multi-way joins, where join execution must be tightly integrated with Datalog's fixpoint semantics.

\paragraph{GPU Databases and Query Processing} systems have extensively studied how to execute relational workloads on heterogeneous CPU-GPU platforms, including column-store execution~\cite{ocelot}, adaptive operator placement across CPUs and GPUs~\cite{adaptive}, and high-performance GPU join implementations~\cite{fastequi}. More recent work has expanded this line to end-to-end optimization of data placement and execution in heterogeneous databases~\cite{orchestrating}, multi-GPU scaling for hybrid DBMSs~\cite{scalinghybrid}, general GPU-accelerated relational execution with load-balancing support~\cite{themis}, improved GPU joins and grouped aggregations~\cite{gustavo_gpujoin}, and execution beyond device memory capacity~\cite{scalinggpu}. These systems substantially advance general-purpose relational analytics on GPU-accelerated database platforms. In contrast, our setting is recursive semi-na\"ive Datalog with worst-case optimal multi-way joins, where join execution must be tightly integrated with iterative delta maintenance, deterministic materialization, and fixpoint semantics, rather than one-shot relational query evaluation.

\section{Conclusion and Future Work}

This paper presented \tool{}, the first WCOJ GPU Datalog engine capable of executing multi-way recursive workloads end-to-end. By fundamentally resolving the intermediate memory exhaustion inherent to exiting binary joins based GPU Datalog system, \tool{} achieves order-of-magnitude speedups over mature CPU baselines beyond simple (\emph{e}.\emph{g}., transitive closure) queries. This robust execution is achieved by balancing the demands of iterative updates and skewed multi-way intersections. Specifically, \tool{} enforces flat columnar storage to guarantee lightweight semi-na\"ive delta merges, utilizes a histogram guide join to deterministically absorb structural skew, and leverages stream parallel rule scheduling to better staturating hardware.

Several directions remain open for future research. At the microarchitectural level, residual inner-variable skew could be mitigated via inter-SM work redistribution on modern GPUs by exploiting architectural features such as Cluster Launch Control. Algorithmically, we plan to develop a GPU-native, cost-model-driven query planner to automatically identify instance-optimal variable orderings. 


\bibliographystyle{ACM-Reference-Format}
\bibliography{sample-base}


\end{document}